\pdfoutput=1
\PassOptionsToPackage{dvipsnames,svgnames,x11names,table}{xcolor}
\RequirePackage{xcolor} 

\documentclass[journal]{vgtc}                     

\onlineid{1999}

\vgtccategory{Research}

\vgtcpapertype{please specify}

\title{Visuo-Tactile Feedback with Hand Outline Styles \\for Modulating Affective Roughness Perception}

\author{%
  \authororcid{Minju Baeck}{0000-0001-7179-2103},
  \authororcid{Yoonseok Shin}{0009-0004-4164-9559}, \authororcid{Dooyoung Kim}{0000-0002-6003-2181}, \authororcid{Hyunjin Lee}{0000-0002-4628-4921}, \authororcid{Sang Ho Yoon*}{0000-0002-3780-5350} and 
  \authororcid{Woontack Woo*}{0000-0002-5501-4421}
}

\authorfooter{
  \item
  	Minju Baeck, Yoonseok Shin is with KAIST UVR Lab.
  	E-mail: \{minjubaeck, sys7498\}@kaist.ac.kr
  \item
  	Dooyoung Kim, Hyunjin Lee is with KAIST KI-ITC ARRC.
  	E-mail: \{dooyoung.kim, clairehj517\}@kaist.ac.kr
  \item
  * Sang Ho Yoon is with KAIST HCI Tech Lab, and Woontack Woo is with KAIST UVR Lab and KAIST KI-ITC ARRC. 
  They are the corresponding authors.
  E-mail: \{sangho, wwoo\}@kaist.ac.kr
}

\abstract{We propose a visuo-tactile feedback method that combines virtual hand visualization and fingertip vibrations to modulate affective roughness perception in VR. While prior work has focused on object-based textures and vibrotactile feedback, the role of visual feedback on virtual hands remains underexplored. Our approach introduces affective visual cues including line shape, motion, and color applied to hand outlines, and examines their influence on both affective responses (arousal, valence) and perceived roughness. Results show that sharp contours enhanced perceived roughness, increased arousal, and reduced valence, intensifying the emotional impact of haptic feedback. In contrast, color affected valence only, with red consistently lowering emotional positivity. These effects were especially noticeable at lower haptic intensities, where visual cues extended affective modulation into mid-level perceptual ranges. Overall, the findings highlight how integrating expressive visual cues with tactile feedback can enrich affective rendering and offer flexible emotional tuning in immersive VR interactions. }


\teaser{
  \centering
  \includegraphics[width=\linewidth, alt={A view of a city with buildings peeking out of the clouds.}]{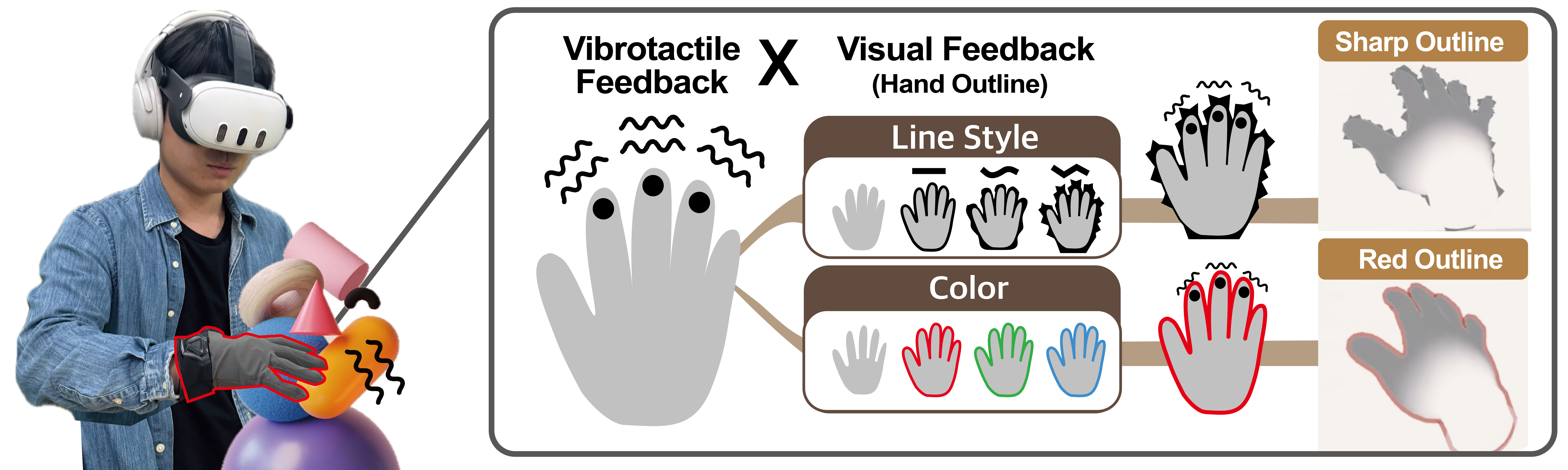}
  \caption{Overview of the visuo-tactile system that modulates affective roughness perception via fingertip vibrations and hand-outline visuals, using \textit{Line style} (\textit{None, Line, Curve–slow/fast, Sharp–slow/fast}) and \textit{Color} (\textit{Red, White, Blue, Green}).}
  \label{fig:teaser}
}

\graphicspath{{figs/}{figures/}{pictures/}{images/}{./}} 

\usepackage{booktabs} 
\usepackage{graphicx}
\usepackage{float} 
\usepackage{makecell} 
\usepackage{xcolor}
\definecolor{darkyellow}{RGB}{204, 204, 0} 
\definecolor{darkgreen}{RGB}{0, 100, 0} 
\definecolor{NavyBlue}{RGB}{0,0,128}
\usepackage{enumitem}
\usepackage{multirow}
\usepackage{comment}
\usepackage{caption}
\usepackage{mathptmx}                  
\usepackage{amsmath}
\usepackage[final,commandnameprefix=ifneeded]{changes}
\usepackage{stfloats} 
\begin{document}


\firstsection{Introduction}

\maketitle

In immersive Virtual Reality (VR), affective responses play a key role in enhancing user experience, engagement, and presence. By incorporating emotional elements, VR can better simulate real-life experiences, impacting areas such as learning, design, and emotional well-being. However, conventional VR feedback relies on straightforward visual cues or direct collision-based vibrations that simply indicate something occurred, lacking the emotional depth necessary for rich and expressive interactions. Valence (pleasant to unpleasant) and arousal (calm to intense) are key affective dimensions~\cite{gunes2013categorical, russell1980circumplex} that facilitate emotionally adaptive interactions in VR. In this study, we use the arousal–valence model proposed by Scherer~\cite{dzedzickis2020human}. Designing affective feedback based on the arousal–valence model enables more nuanced, emotionally resonant, and context-aware experiences in VR, enhancing immersion and allowing systems to adapt to users' emotional states. 

Emotional responses play a crucial role in interactions with objects, particularly with tactile feedback. Roughness is a key perceptual attribute in tactile texture perception~\cite{gescheider2005perception, hollins2007coding}, and emotional responses to roughness naturally emerge~\cite{drewing2017feeling}. Fingertip vibration is highly sensitive~\cite{gonzalez2014analysis}, and research on vibrotactile roughness perception has been actively explored~\cite{asano2014vibrotactile,hollins2001vibrotactile}. 
Humans process information through multiple sensory channels, with visual and tactile stimuli often interacting to enhance or complement each other~\cite{ernst2002humans, guest2003role}. Previous research on roughness perception has discovered that visual feedback can modulate the perceived roughness perception~\cite{suzuishi2020visual, ujitoko2019modulating}. 

In affective computing, visual parameters such as color, shape, and motion are known to impact emotional perception~\cite{lockyer2012affective, omata2012affective, wilson2017multi}. For instance, red can induce arousal and urgency, while blue or green evoke calmness or safety~\cite{kaya2004relationship, suk2008emotional}. While prior work has investigated the emotional effects of visual cues in general VR environments (e.g., pain modulation~\cite{martini2013color}), the role of visual feedback in shaping affective roughness perception remains underexplored. 

\added{Despite these insights, design guidelines for emotion-inducing virtual environments are still limited~\cite{dozio2022design}. However, prior research has shown that emotions can be effectively evoked by combining multiple sensory modalities, such as vision and touch. While these modalities can be perceived independently, their combination may produce emergent or synergistic effects that are not predictable from each modality alone. Designing multimodal feedback is essential not only for enhancing emotional immersion and perceptual consistency, but also for compensating for limitations in resource-constrained VR settings. This motivates the need for systematic investigations into how visual and tactile cues interact to shape emotional experience in VR environments.}

We propose a visuo-tactile feedback method that integrates expressive hand-centered visual cues with fingertip vibrotactile feedback to modulate affective roughness perception during hand–object interaction (see ~\cref{fig:teaser}). Specifically, we designed visual feedback cues that directly map to valence and arousal by modifying hand outline properties in real time. We focused on two affective visual parameters: \textit{Line Style} (shape and motion) and \textit{Color}, which were overlaid on the virtual hand during the interaction. \textit{Line style} represents waveform sharpness and movement dynamics, while \textit{Color} reflects affective associations grounded in prior studies.

We hypothesized that such hand-centered affective visual feedback combined with vibrotactile stimuli could enhance or modulate users’ emotional responses to roughness beyond what tactile feedback alone can provide. To validate this, we conducted two user studies: 1)~a preliminary study to identify vibrotactile patterns representing distinct roughness levels and their affective associations, and 2) a main study examining how \textit{Line Style} and \textit{Color}, when paired with haptic stimuli, influence affective responses (valence, arousal) and perceived roughness. This approach aims to enrich affective and sensory experiences in VR through visuo-tactile integration.

To guide this investigation, we pose the following research questions:
\begin{itemize}
    \item \textbf{RQ1.} How do \textit{Line Style} and vibrotactile feedback shape affective responses (valence, arousal) and perceived roughness?
    \item \textbf{RQ2.} How does \textit{Color} influence affective responses (valence, arousal) and roughness perception when paired with haptic feedback?
\end{itemize}
\vspace{- 0.4em}
Overall, our main contributions are as follows. First, we propose a visuo-tactile feedback method that combines virtual hand outline visualization with fingertip vibrations to modulate affective roughness perception during hand–object interaction in VR, focusing on valence and arousal. This approach shifts the focus from object-centered texture rendering to hand-centered visual feedback, allowing for more expressive emotional modulation. Second, our empirical findings show that both \textit{Line styles} and vibrotactile stimuli significantly influence perceived roughness, arousal, and valence, while \textit{Color} primarily affects valence. These effects are amplified when visual cues are integrated, compared to tactile-only feedback. Third, we offer design insights on how expressive hand-outline cues can be used to deliver affective feedback, guiding the development of emotionally adaptive multimodal feedback systems in immersive environments.

\section{Related work}
\subsection{Immersive Affective VR Experience}

VR is increasingly recognized as a powerful medium for inducing and assessing rich emotional experiences, particularly within the framework of appraisal theory~\cite{meuleman2018induction}. This is largely due to VR’s ability to generate a strong sense of presence, allowing users to safely experience emotions such as anxiety and relaxation in controlled environments~\cite{riva2007affective}. To effectively support such affective interactions, the Circumplex Model of Affect is commonly used to evaluate emotional states along two key dimensions: arousal and valence~\cite{russell1980circumplex}. 

Affective computing allows VR to adjust to how users feel, helping to keep them more engaged and making the overall experience more effective~\cite{marin2020emotion}. Since VR is no longer just about watching but also about interacting, it's more important than ever to design emotional experiences that happen during those interactions. Such affective response design plays a critical role in narrative or task-driven VR scenarios, ranging from stress-inducing training~\cite{talamo2023impact}, to therapeutic relaxation~\cite{prabhu2019affective}, and immersive experiences such as entertainment and virtual tourism~\cite{jung2024hapmotion,bayro2024advancing}. Although previous research has examined emotional responses to passive touch or simply viewing virtual environments and social contexts~\cite{gabana2017effects, prabhu2019affective, sykownik2020experience}, affective experiences during active object exploration in virtual environments remain largely underexplored. This study investigates how visual modulation affects affective responses when vibrotactile roughness is presented in object–hand interactions.

\subsection{Vibrotactile Feedback for Roughness and Emotion}
Vibrotactile cues, particularly variations in frequency, intensity, and waveform, have been widely used to simulate different levels of surface texture~\cite{ahmaniemi2010design, asano2014vibrotactile}. High-frequency or irregular vibrations are typically associated with coarse textures, while low-frequency, smooth vibrations represent softer or smoother surfaces~\cite{hollins2001vibrotactile}. Previous research has explored how vibrotactile feedback can be mapped to the perceived roughness of textured surfaces, aiming to simulate tactile sensations based on texture information.

Beyond texture perception, affective haptics has explored how vibration parameters influence emotional experiences such as arousal and valence~\cite{akshita2015towards, wilson2017multi, yoo2015emotional}. For instance, increasing the perceived roughness of vibrotactile textures tends to elicit higher arousal and lower valence~\cite{seifi2013first, yoo2015emotional}. In contrast to traditional unimodal haptic approaches, this study investigates how augmenting fingertip vibrotactile feedback with visual cues affects both emotional responses and roughness perception, aiming to expand the expressive potential of tactile interaction through cross-modal enhancement.

\subsection{Crossmodal Visuo-Tactile Interaction}
Visual feedback has been shown to modulate roughness perception. For instance, pseudo-haptic effects from oscillating visual pointers enhanced perceived roughness even without changes in tactile input~\cite{ujitoko2019modulating}, and task-irrelevant visual motion was found to influence roughness during active touch~\cite{suzuishi2020visual}. Vibrotactile stimuli can crossmodally associate with visual features such as color and contour. For instance, high roughness vibrations are frequently perceived as warm and dark in color and are associated with heightened arousal and pleasure~\cite{neto2025investigating}, highlighting the close relationship between touch, vision, and emotion. 

In affective computing, visual parameters such as color~\cite{rajae2011effects, wilms2018color}, motion~\cite{lockyer2012affective}, and shape (e.g., curvature or angularity)~\cite{lockyer2012affective, valtchanov2015enviropulse} are widely used to convey emotional tone. Curved visuals often signal calmness or positivity, while sharp or erratic forms suggest tension or urgency~\cite{lockyer2012affective}. Colors are consistently mapped to levels of valence and arousal~\cite{suk2008emotional, zhang2019visualtouch}. Visual representations in immersive environments can modulate both tactile and emotional perception by influencing how users align sensory input with their virtual body. For example, even skin color can alter pain sensitivity, such as red lowering heat pain thresholds~\cite{martini2013color}. This highlights the power of visual cues in shaping experience and presence in VR~\cite{normand2024different, sanchez2005presence}. Moreover, visual features like color and motion can directly evoke emotion~\cite{valtchanov2015enviropulse}, and their combination with haptic cues expands the affective design space~\cite{wilson2017multi}.

These findings suggest that visual cues applied to the virtual hand could similarly influence how users perceive and emotionally respond to tactile experiences. Building on this, our study investigates whether affective visual cues on the hand can modulate emotional roughness perception when paired with fingertip vibrotactile input.

\section{Method}
This study proposes a method to modulate emotional roughness perception by combining fingertip vibrotactile patterns with visual feedback, such as \textit{Line Style} and \textit{Color}. We examine how visual hand cues influence affective roughness perception in VR. 

\subsection{Tactile Feedback Design for Roughness}
This section details the design of a vibrotactile system using the TactGlove DK2~\cite{bHapticsTactGlove}, aimed at simulating different levels of roughness perception during hand-object interaction in VR. This glove supports linear resonant actuator (LRA) based vibrotactile feedback and hand tracking, which allows natural hand-object interaction while ensuring accurate feedback in various settings. 

\replaced{Previous work showed that roughness perception increases with higher vibrotactile intensity and vibratory energy~\cite{hollins2000imposed}. Here, we adjusted both stimulus intensity and on-duration to generate distinguishable levels of perceived roughness, and validated whether these differences were perceptible through the TactGlove DK2.}{Previous research suggests that higher intensity and shorter duration are typically associated with stronger sensations of roughness~\cite{hollins2000imposed}. To create distinguishable roughness levels, we adjusted both the intensity and duration of vibrotactile stimuli} According to manufacturer specifications, the TactGlove DK2 uses LRAs with a resonant frequency of $170\,\text{Hz}$ and delivers approximately $1\,\text{G}$ peak acceleration at 100\% intensity. Based on \replaced{these constraints,}{this range,} we selected two intensity levels\deleted{for the experiment}: \replaced{10\% (low), approximating a light mobile vibration, and 80\% (high), a strong smartwatch alert, were chosen to reflect a perceptual range of vibrotactile strength.}{10\% (low), mimicking a subtle mobile vibration, and 80\% (high), resembling a strong smartwatch alert.} To \replaced{accommodate}{fit} the  $40\,\mathrm{ms}$ timing constraint of the TactGlove DK2, \replaced{we varied the on-duration across four values}{used four pulse durations}—$5\,\mathrm{ms}$, $20\,\mathrm{ms}$, $30\,\mathrm{ms}$, and $40\,\mathrm{ms}$—\replaced{to modulate the total vibratory exposure per cycle.}{within a fixed $40\,\mathrm{ms}$ cycle, varying only the vibration-on time while maintaining a consistent rhythm. }

\replaced{To simulate real-world tactile sensations during hand–object contact, vibrotactile stimuli were delivered simultaneously to the index, middle, and ring fingers. }{To enhance immersion, vibrotactile stimuli were simultaneously applied to the index, middle, and ring fingers, simulating surface contact and reflecting how tactile sensations are perceived in real-world hand–object interactions. }\replaced{Eight vibrotactile conditions were created by combining two intensity levels with four durations.}{Eight vibrotactile conditions were generated by combining the two intensity levels with the four duration settings.} \replaced{To verify whether these parameters could reliably convey differences in perceived roughness, we conducted a preliminary user study. (see \cref{chap:preliminary}). Results confirmed that participants could clearly distinguish roughness levels across these stimuli, confirming their suitability for the main study.}{These conditions served as candidate stimuli for a preliminary study, the details of which are presented in } \cref{chap:preliminary}.

\subsection{Affective Visual Feedback Design}
To modulate affective perception of roughness during hand–object interaction in VR, we introduced animated hand outlines that deliver emotional cues without interfering with task performance. Rather than modifying the hand's shape or surface texture, we adopted a less obstructive, transparent black virtual hand~\cite{voisard2023effects}, and focused on modulating its outline using \textit{Line Style} and \textit{Color}.

\begin{figure}[t]
  \centering
  \includegraphics[width=\columnwidth]{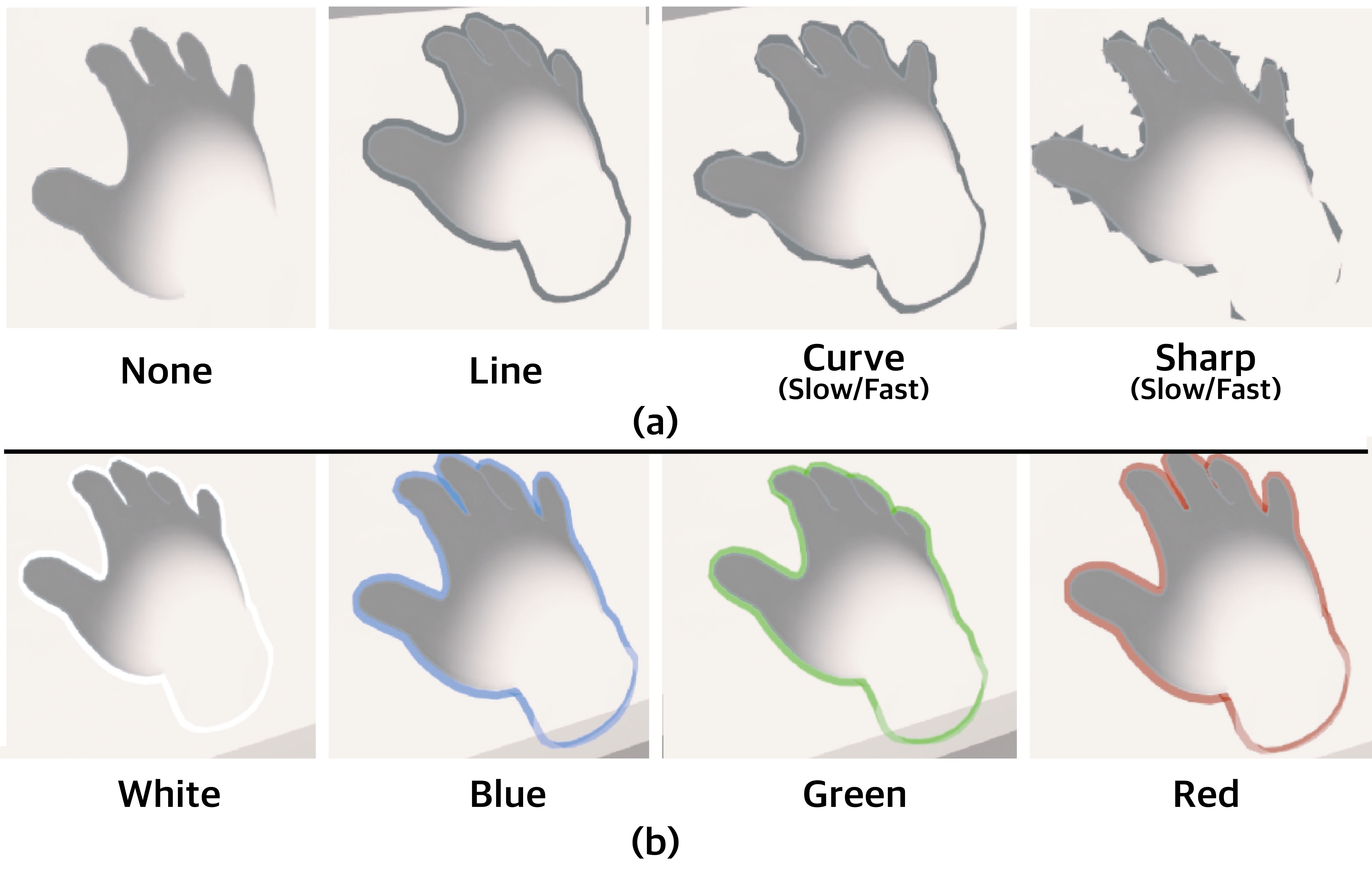}
  \caption{Visual feedback stimuli. 
(a) \textbf{Line Style}: None, Line, Curve (Slow/Fast), Sharp (Slow/Fast). 
(b) \textbf{Color}: Static Line Style with White, Blue, Green, or Red outlines.}
  \label{fig:flowchart2222}
  \vspace{-0.4cm}
\end{figure}

This approach builds on findings that visual fidelity is not essential for agency or ownership~\cite{argelaguet2016role, kilteni2012sense}, and on prior research showing that motion, curvature, and color can shape emotional perception~\cite{lockyer2012affective, rajae2011effects, suk2008emotional, valtchanov2015enviropulse, wilms2018color}. We therefore designed two core visual parameters: \textit{Line Style}, which combines shape and motion; and \textit{Color}, applied independently. This separation enables both individual and combined evaluation of how dynamic outlines and chromatic cues influence affective roughness.  Specifically, we adapted the Quick Outline Unity asset\footnote{\url{https://assetstore.unity.com/packages/tools/particles-effects/quick-outline-115488}} to support real-time modulation synchronized with fingertip vibrotactile feedback.  \added{An overview of the visual stimuli used in the study is presented in Fig.~\ref{fig:flowchart2222}.}

\subsubsection{Design of Affective \textit{Line Style} Feedback}
We defined hand outline styles based on the combination of spatial form (\textit{None}, \textit{Line}, \textit{Curve}, \textit{Sharp}) and temporal motion (\textit{Static}, \textit{Slow}, \textit{Fast}). \textit{Line style} was implemented by modifying the position of each vertex as a function of the time-varying parameter $t$, resulting in dynamic deformations synchronized with the animation timeline.

\vspace{-0.5cm}
\begin{gather}
t = t_e  s_w
\end{gather}

where $t_e$ represents the accumulated elapsed time since the beginning of the unity scene. $s_w$ denotes the wave speed, determining the temporal frequency and dynamism of the animated outline.
To ensure perceptual clarity and consistent affective expression across visual styles, wave speeds were empirically calibrated through a pilot study. Different \textit{Line style}s required different speed values due to the inherent temporal characteristics of their deformation functions. Therefore, $s_w$ was tuned per style to match perceptual salience and emotional impression, rather than applying a uniform value across conditions. Each \textit{Line Style} is described below to clarify its visual and temporal properties.

\textbf{None} No visual outline was rendered. The None condition served as a baseline to assess tactile feedback in isolation. This baseline condition allowed us to evaluate tactile feedback without visual influence.

\textbf{Line} This style represents a static outline with no temporal modulation. It is implemented by setting \( s_w = 0.0 \), effectively disabling all time-dependent deformation. This condition serves as a baseline to isolate the visual impact of outline presence and color without introducing motion cues.

\textbf{Curve} 
A smooth, pulsating outline was implemented using a smoothed value noise function to produce continuous, organic deformation. The noise is defined as:
\begin{multline}
\text{curveNoise}(x) \\=  (1 - s(\text{frac}(x))) \cdot h(\lfloor x \rfloor)  + s(\text{frac}(x)) \cdot h(\lfloor x \rfloor + 1)
\end{multline}
where $h(x)$ is a pseudo-random hash function, $s(x)$ is a smoothing interpolation, and $\text{frac}(x)$ extracts the fractional part of $x$. This generates smoothly varying noise between points.

The deformation offset is calculated as:
\begin{gather}
x' = p_{x}f + t,\quad y' = 1.2p_{y}f + t \\
o_{curve} = \frac{1}{2} S(\text{curveNoise}(x')^2 + \text{curveNoise}(y')^2)
\end{gather}
where $p_x$, $p_y$ are vertex positions, $f$ is wave frequency, and $S$ is a scaling factor. Values are squared to enhance dominant features, and a slight vertical multiplier introduces asymmetry for more natural motion. This outline was designed to suggest pleasant or soft roughness. To differentiate emotional intensity, we implemented two variants: \textit{Slow} and \textit{Fast}, using distinct wave speeds of $s_w = 2.0$ and $s_w = 10.0$, respectively.

\textbf{Sharp} 
A jagged, fluctuating outline was generated using a sawtooth waveform:
\begin{equation}
\text{sharpNoise}(x) = 2(\text{frac}(x) - 0.5)
\end{equation}
which produces a repeating ramp from $-1$ to $1$.

The deformation offset is:
\begin{gather}
x' = p_{x}f + t,\quad y' = p_{y}f + t \\
o_{sharp} = \frac{1}{2} S(\text{sharpNoise}(x') + \text{sharpNoise}(y'))
\end{gather}
This style visually conveys tension or discomfort, aligning with tactile impressions of roughness. As with the \textit{Curve} style, we defined two versions: \textit{Slow} and \textit{Fast}, using different wave speeds of $s_w = 0.25$ and $s_w = 2.0$ to modulate the temporal intensity of deformation.

\subsubsection{ Design of Affective \textit{Color} Feedback} To incorporate affective modulation into tactile experiences, we selected \textit{Red, Green, Blue}, and \textit{White} based on prior research demonstrating their significant effects on valence and arousal~\cite{suk2008emotional, zhang2019visualtouch}. \textit{Red} has been shown to evoke negative valence and high arousal, often associated with urgency or irritation. In contrast, \textit{Green} and \textit{White} are typically linked to higher valence and moderate arousal, conveying feelings of safety, balance, and neutrality. \textit{Blue} is generally perceived as calming and pleasant, associated with positive valence and lower arousal. These affective \textit{color} properties were used to guide users' emotional interpretation of tactile roughness, particularly when overlaid on hand outlines during interaction. An overview of the corresponding visual conditions and parameters is provided in~\cref{tab:visual_conditions}.

\begin{table}[H]
    \centering
    \caption{\added{Stimulus types determined by combinations of \textit{Line Style} and \textit{Color} variables. \textit{Line Style} varies by waveform and speed in gray, and \textit{Color} is applied to the static \textit{Line} style.}}
    \begin{tabular}{cccc}
    \toprule
        \textbf{Visual} & \textbf{Line Style} & \textbf{$s_w$} & \textbf{Color} \\
    \midrule
        \multirow{4}{*}{\shortstack{Line Style \\ (6 types)}} 
        & \textit{None}          & --    & \multirow{4}{*}{\textit{Gray}} \\
        & \textit{Line}          & 0.0   & ~ \\
        & \textit{Curve\_(Slow/Fast)}   & 2.0 / 10.0  & ~ \\
        & \textit{Sharp\_(Slow/Fast)}   & 0.25 / 2.0 & ~ \\
    \midrule
        \multirow{2}{*}{\shortstack{Color \\ (4 types)}} &
        \multirow{2}{*}{\textit{Line}} &
        \multirow{2}{*}{0.0} 
         & \textit{Red, Green} \\
        &                           &       & \textit{Blue, White} \\
    \bottomrule
    \end{tabular}
    \deleted{Table 1: Stimulus types determined by combinations of \textit{Line Style} and \textit{Color} variables. \textit{Line Style} varies by waveform and speed in gray, and \textit{Color} is applied to the static \textit{Line} style.}
    \label{tab:visual_conditions}
\end{table}

\section{Preliminary Study: Haptic Roughness}
\label{chap:preliminary}
\replaced{This chapter presents a preliminary study aimed at identifying vibrotactile stimuli that evoke reliably distinguishable levels of roughness. The study empirically validates the parameter design introduced in Chapter 3 and selects representative low, medium, and high roughness conditions for the main study.}{This chapter presents a preliminary study conducted to identify a subset of vibrotactile stimuli that evoke clearly distinguishable levels of roughness perception. The goal was to empirically validate the design parameters introduced in Chapter 3 and select representative low, medium, and high roughness stimuli for use in the main study.}

\begin{figure}[b]
  \centering
  \includegraphics[width=\columnwidth]{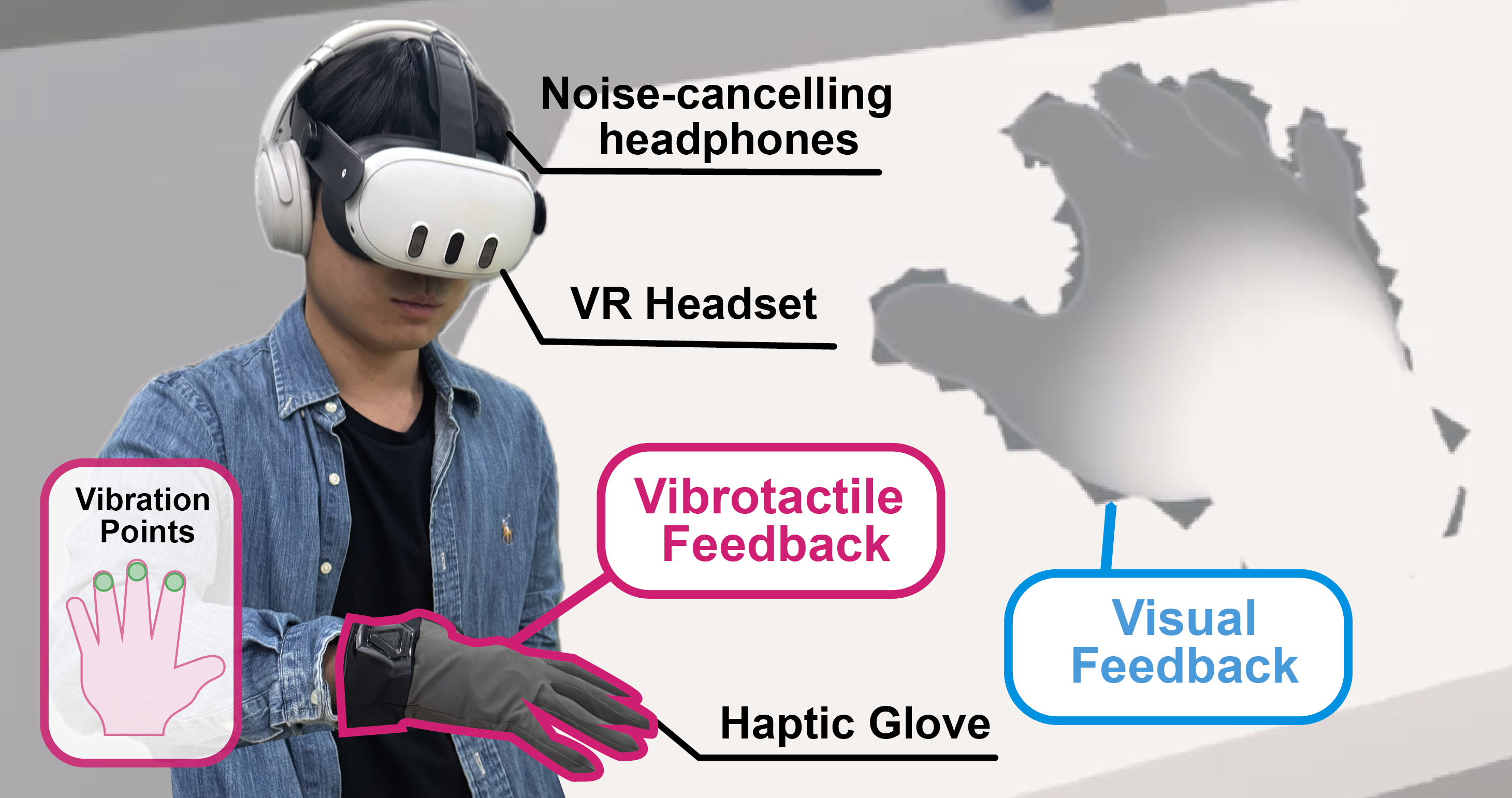}
  \caption{System setup showing visual feedback (animated sharp hand outlines) and vibrotactile feedback during hand–object interaction using VR headset and haptic glove.}
  \label{fig:flowchart2}
\vspace{-0.3cm}
\end{figure}

\subsection{Apparatus}
The experiment used the TactGlove DK2 by bHaptics, which delivers fingertip vibrotactile feedback through LRA motors. Stimuli were designed and calibrated using the bHaptics Designer tool. The haptic glove supported hand tracking and communicated via Bluetooth. The VR environment was built in Unity and operated on a Meta Quest 3 headset.
~\cref{fig:flowchart2} illustrates the system setup, where visual hand outline animation and vibrotactile feedback via TactGlove DK2 are triggered upon virtual hand–object contact using Meta Quest 3. Feedback persisted for up to 3 seconds or until the hand left the object, at which point all feedback was deactivated.
\subsection{Participants and Procedure}
A total of 10 participants (six male, four female), aged between 24 and 30 years (M = 28.1), took part in the study. All were right-handed, with limited experience using haptic gloves (fewer than five times). Two had no prior experience with VR devices.

Participants completed two blocks, each containing all eight vibrotactile stimuli in a randomized order. A one-minute break was provided between blocks. Each stimulus triggered a vibration lasting up to three seconds when participants touched a virtual object with their palm. A moving bar guided hand movement at a constant speed of 0.17~m/s, and participants could repeat the interaction as needed. To mask glove motor noise, Bose QC noise-canceling headphones playing white noise were used. No visual feedback was provided—only haptic sensations were evaluated. After perceiving each stimulus, participants rated arousal and valence (on a 9-point scale) and perceived roughness (1–100) using an on-screen slider interface. Affective responses were measured using the Self-Assessment Manikin (SAM)~\cite{bradley1994measuring}, which uses pictorial 9-point Likert scales (1 = low/negative, 9 = high/positive) for arousal and valence. The session lasted approximately 10 minutes. The study was approved by the IRB, and all participants provided informed consent. No monetary compensation was provided.

\begin{figure}[t]
\centering
\includegraphics[width=\columnwidth]{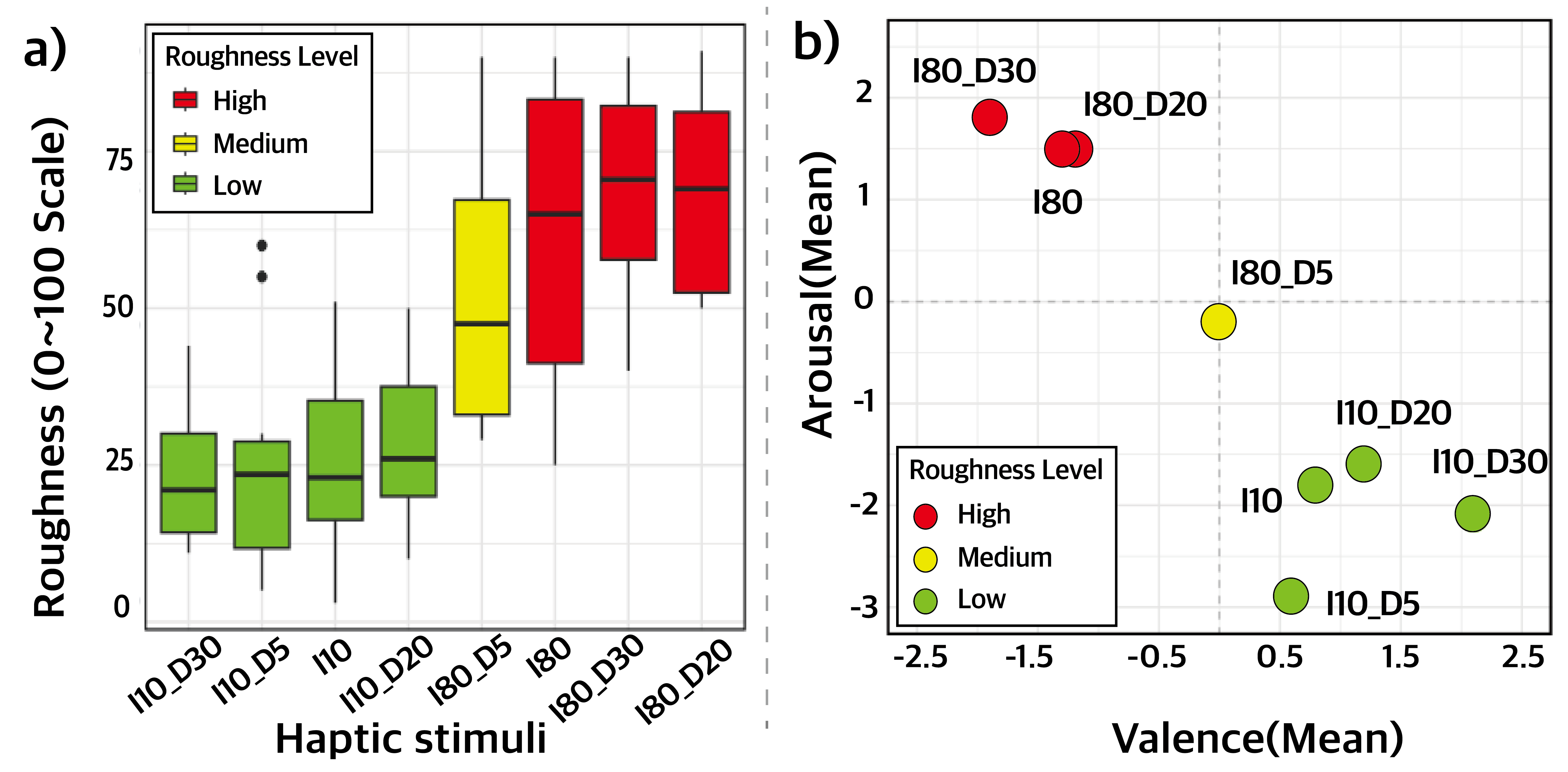}
\caption{
(a) Roughness scores across haptic stimuli.
(b) Valence–arousal plot for haptic stimuli, color-coded by roughness level.}
\label{fig:s1result}
\vspace{-0.3cm}
\end{figure}

\subsection{Results}
We excluded the first trial from analysis to minimize potential effects of unfamiliarity or initial adjustment. We used data from the second trial for all analyses. ~\cref{fig:s1result}  shows that higher roughness levels are associated with increased arousal and decreased valence.

\textbf{Perceived roughness.} The eight vibrotactile stimuli were categorized into three levels of perceived roughness—\textit{Low}, \textit{Medium}, and \textit{High}—based on equal-interval binning of their mean roughness ratings. This classification was used to analyze how visual feedback modulates tactile perception across distinct haptic intensities.
\cref{tab:roughness_valence_arousal} summarizes the mean and standard deviation of roughness, valence, and arousal ratings for each condition.

\textbf{Arousal.} The mean arousal scores varied across haptic conditions. The high roughness conditions showed high arousal values, with I80\_D30 (M = 1.8, SD = 1.40), I80\_D20 (M = 1.5, SD = 1.58), and I80 (M = 1.5, SD = 1.58). 
In contrast, the low roughness conditions resulted in lower arousal, including I10\_D30 (M = -2.1, SD = 1.52), I10\_D5 (M = -2.9, SD = 0.57), I10\_D20 (M = -1.6, SD = 1.58), and I10 (M = -1.8, SD = 1.23). 
The Medium roughness condition (I80\_D5) showed a near-neutral arousal level (M = -0.2, SD = 1.32).

\textbf{Valence.} 
Valence scores were generally higher for Low roughness stimuli, such as I10\_D30 (M = 2.1, SD = 1.79), I10\_D20 (M = 1.2, SD = 2.04), I10 (M = 0.8, SD = 1.69), and I10\_D5 (M = 0.6, SD = 2.37). 
In contrast, High roughness stimuli exhibited lower valence, including I80\_D30 (M = -1.9, SD = 1.45), I80\_D20 (M = -1.2, SD = 1.81), and I80 (M = -1.3, SD = 1.42). 
The Medium roughness condition (I80\_D5) produced a neutral valence score (M = 0.0, SD = 1.33).

\begin{table}[b]
\centering
\caption{Mean (SD) of Roughness (R), Valence (V), and Arousal (A) across each Haptic Condition, and Valence–Arousal Profile}
\resizebox{\columnwidth}{!}{%
\begin{tabular}{lccccp{2.6cm}}
\toprule
\textbf{Haptic} & \textbf{R Mean (SD)}  & \textbf{V Mean (SD)} & \textbf{A Mean (SD)} & \textbf{V–A Profile} \\
\midrule
\textcolor{red}{I80\_D20} & 69.0 (16.2) & -1.2 (1.81) & 1.5 (1.58) & Negative V, High A \\
\textcolor{red}{I80\_D30}  & 68.7 (17.0) & -1.9 (1.45) & 1.8 (1.40) & Negative V, High A \\
\textcolor{red}{I80}& 61.1 (7.84) & -1.3 (24.8)  & 1.5 (1.58) & Negative V, High A \\
\textcolor{darkyellow}{I80\_D5}& 52.2 (21.3) & 0.0 (1.33) & -0.2 (1.32) & Neutral V, Neutral A \\
\textcolor{darkgreen}{I10\_D\replaced{2}{1}0} & 27.9 (12.8)& 1.2 (2.04) & -1.6 (1.58) & Positive V, Low A \\
\textcolor{darkgreen}{I10\_D30} & 26.2 (16.0)  & 2.1 (1.79) & -2.1 (1.52) & Positive V, Low A \\
\textcolor{darkgreen}{I10\_D5} & 25.4 (19.0) & 0.6 (2.37) & -2.9 (0.57) & Positive V, Low A \\
\textcolor{darkgreen}{I10} & 22.8 (10.9) & 0.8 (1.69) & -1.8 (1.23) & Positive V, Low A \\
\bottomrule
\end{tabular}}
\label{tab:roughness_valence_arousal}
\small
\textit{Note.} Color : roughness (green: low, yellow: medium, red: high).  I10/I80 = 10\%/80\% intensity; D5–D30 = 5–30~ms duration; no suffix = 40~ms. V–A Profile = mean valence and arousal classification.
\vspace{-0.4cm}
\end{table}

\subsection{Discussion}

The preliminary study identified a set of vibrotactile stimuli that evoke distinguishable levels of perceived roughness, while also revealing meaningful differences in affective responses. The classification into \textit{Low}, \textit{Medium}, and \textit{High} roughness based on mean ratings allowed us to select representative stimuli for the main study.

High-intensity, short-duration conditions (e.g., I80\_D30, I80\_D20) were consistently perceived as rougher and evoked higher arousal but lower valence, suggesting they were interpreted as more intense and less pleasant. This trend aligns with previous research indicating that higher levels of perceived roughness are often associated with more negative emotional responses. Conversely, low roughness  stimuli (e.g., I10\_D30, I10\_D20) were associated with lower roughness and arousal, but higher valence, indicating a more subtle and pleasant tactile experience. 

Interestingly, one mid-level stimulus (I80\_D5) yielded neutral scores for both arousal and valence, supporting its suitability as a medium-roughness reference in the subsequent study. These findings align with prior research showing that increased vibrotactile intensity and temporal sharpness enhance perceived roughness and emotional arousal~\cite{shetty2021emotional, yoo2015emotional}. The absence of visual cues in this study ensured that the reported effects were driven purely by tactile stimulation, establishing a controlled foundation for assessing crossmodal modulation in the main study. We selected three stimuli with clearly different levels of roughness and affective response and used them as the \textit{Low} (I10\_D30), \textit{Medium} (I80\_D5), and \textit{High (I80\_D30)} haptic conditions in the visuo-tactile integration experiment.

\section{Main study: Visuo-Tactile Feedback} %

This study investigates whether the integration of vibrotactile and affective visual cues (\textit{Line Style} and \textit{Color}) influences affective responses (valence, arousal) as well as perceived roughness.

\subsection{Stimuli}
We examined the effects of two different conditions (\textit{Line Style}, \textit{Color}) in combination with three levels of vibrotactile roughness (\textit{Low}, \textit{Medium}, \textit{High}), selected from our preliminary study based on perceived roughness and affective profiles (see ~\cref{tab:haptic1}).
\vspace{-0.3em}
\begin{table}[h]
\centering
\caption{Selected Haptic Stimuli and Valence–Arousal Profiles}
\label{tab:haptic1}
\resizebox{0.95\linewidth}{!}{%
\begin{tabular}{lll}
\toprule
\textbf{Roughness Level} & \textbf{Stimulus} & \textbf{Valence–Arousal Profile} \\
\midrule
RL (Low)     & \textit{I10\_D30} & Positive valence, Low arousal \\
RM (Medium)  & \textit{I80\_D5} & Neutral valence, Neutral arousal \\
RH (High)    & \textit{I80\_D30} & Negative valence, High arousal \\
\bottomrule
\end{tabular}
}
\vspace{-1em}
\end{table}

\textit{Line Style} included six levels (e.g., \textit{None}, \textit{Line}, \textit{Curve}, \textit{Sharp}, each with slow/fast variants), resulting in 18 conditions (3 \textit{Haptic} $\times$ 6 \textit{Line Style}). \textit{Color}  included four levels (\textit{Red, Blue, Green, White}), applied only to the \textit{Line} condition, forming 12 conditions (3 \textit{Haptic} $\times$ 4 \textit{Color}). 

To systematically examine the contribution of each visual factor, the study was divided into two visual conditions.
. The \textit{Line Style} condition tested the combination of \textit{Line Style} and \textit{Haptic}  (18 stimuli), while the \textit{Color} condition tested the combination of \textit{Color} and \textit{Haptic} (12 stimuli). 
For clarity, we refer to the selected haptic stimuli using the following abbreviations:  
\textbf{RH} (Roughness High), \textbf{RM} (Roughness Medium), and \textbf{RL} (Roughness Low).

\subsection{Participants and Procedure}

We omitted the methods common to the preliminary study for brevity. A total of 24 participants (12 male, 12 female; age 19–34, M = 25.2) took part in the study. All but one were right-handed. Ten had prior VR experience, and all but one had limited exposure to haptic gloves (fewer than five uses). None had participated in the preliminary study.

\begin{figure*}[t]
 \centering
 \includegraphics[width=\textwidth]{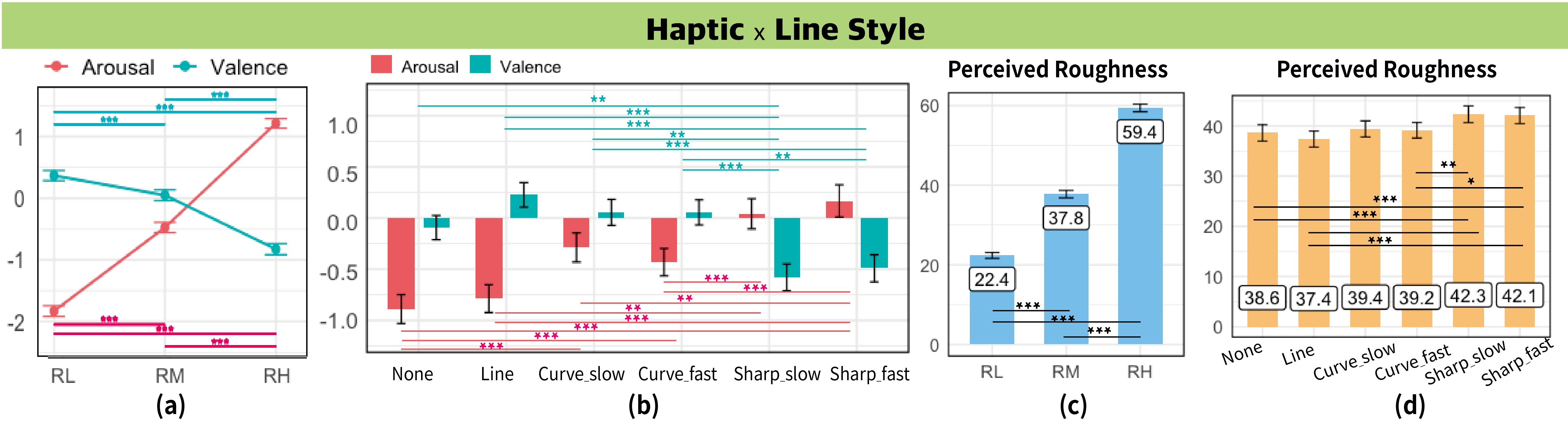}
\caption{Effects of \textit{Haptic} and \textit{Line Style} on affective responses and perceived roughness: 
(a) Line plot showing mean arousal and valence across \textit{Haptic} conditions (RL, RM, RH). 
(b) Box plots of arousal and valence across \textit{Line Style}. 
(c) Mean perceived roughness by \textit{Haptic}. 
(d) Mean perceived roughness by \textit{Line Style}. 
Significance: *$p<.05$, **$p<.01$, ***$p<.001$.}
    \label{fig:line_col222222}
\vspace{-0.1cm}
\end{figure*}

All participants completed both visual conditions: the \textit{Line Style} condition and the \textit{Color} condition, each consisting of three trials. In the  \textit{Line Style}  condition, participants were presented with 18 stimuli (3 \textit{Haptic} $\times$ 6 \textit{Line Styles}) per trial; in the \textit{Colors} condition, 12 stimuli (3 \textit{Haptic} $\times$ 4 \textit{Color}) were shown per trial. The stimulus order was randomized for each trial and participant. The order of the two visual conditions was counterbalanced across participants to minimize order effects. Each condition was completed in full before proceeding to the next, and a one-minute rest was provided between trials. The study protocol was approved by the institutional review board, and all participants provided informed consent. Each trial took approximately 4 minutes, and the entire experiment lasted about 45 minutes. Each participant was compensated with approximately USD 10.

\subsection{Apparatus}
We used the same apparatus and system setup as in the preliminary study, including the TactGlove DK2 for vibrotactile feedback and a Unity-based VR environment on the Meta Quest 3 (see ~\cref{fig:flowchart2}).

The interaction procedure followed the same setup as in the preliminary study: participants stroked a virtual object with their palm, triggering up to 3 seconds of vibrotactile feedback. A moving bar guided their hand at a constant speed ($.17\,\mathrm{m/s}$), and participants could repeat the interaction freely. To mask glove motor noise, participants wore noise-canceling headphones playing white noise. After each stimulus, participants rated arousal (1–9), valence (1–9), and perceived roughness (1–100) using an on-screen slider. Affective responses were measured using the Self-Assessment Manikin (SAM)~\cite{bradley1994measuring}.

We analyzed statistical data using RStudio(version 4.4.1). The Shapiro–Wilk test was performed to assess the normality of the data. For variables that violated normality assumptions, we employed the aligned rank transform (ART) procedure using the ARTool package~\cite{wobbrock2011aligned}. Repeated-measures ANOVAs were conducted to examine main and interaction effects with a significance threshold of $\alpha = .05$. For significant main or interaction effects, post hoc pairwise comparisons were conducted using Bonferroni-adjusted $t$-tests for both normally distributed and ART-transformed data. For ART-transformed data, the pairwise $t$-tests were performed on estimated marginal means (EMMs) via the emmeans package. 


\subsubsection{Condition I: \textit{Haptic} $\times$ \textit{Line Style}}
The Shapiro–Wilk test indicated significant deviations from normality (all $p < .05$) for arousal, valence, and roughness scores, prompting the use of non-parametric ART ANOVA. ~\cref{fig:line_col222222} visualizes the main effects of \textit{Haptic} roughness and \textit{Line Style} on average arousal, valence, and perceived roughness.

\textbf{Valence.} Valence was significantly reduced by both high roughness haptic stimuli and sharp line styles, indicating that these cues independently contributed to negative affective responses. 
The results revealed significant main effects of \textit{Haptic} ($F_{2,1255} = 62.35$, $p < .001$, $\eta^2_p = .09$) and \textit{Line style} ($F_{5,1255} = 10.50$, $p < .001$, $\eta^2_p = .04$), but no significant interaction effect between the two factors ($F_{10,1255} = .41$, $p = .943$, $\eta^2_p < .001$). Post-hoc comparisons were conducted independently for each main effect using pairwise $t$-tests with Bonferroni correction: one for the \textit{Haptic} conditions and one for the \textit{Line Style} conditions. The high roughness haptic condition (\textit{RH}; $M = -.83$, $SD = 1.88$) elicited significantly lower valence ratings than both the low roughness  (\textit{RL}; $M = .37$, $SD = 1.70$, $t(1255) = -10.94$, $p < .001$, $d = .62$) and medium roughness conditions (\textit{RM}; $M = .05$, $SD = 1.80$, $t(1255) = -7.41$, $p < .001$, $d = .42$). The \textit{RL} condition also led to significantly higher valence than the \textit{RM} condition ($t(1255) = 3.53$, $p < .01$, $d =.20$). For the \textit{Line Style} conditions, both \textit{sharp\_slow} ($M = -.58$, $SD = 1.92$) and \textit{sharp\_fast} ($M = -.49$, $SD = 1.95$) elicited significantly lower valence than \textit{curve\_slow} ($M = .06$, $SD = 1.87$), \textit{curve\_fast} ($M = .06$, $SD = 1.81$), and \textit{line}($M = .23$, $SD = 1.76$). For \textit{sharp\_slow} , this difference was significant for \textit{curve\_fast} ($t(1255) = 4.43$, $p < .001$, $d = .25$), \textit{curve\_slow} ($t(1255) = 4.28$, $p < .001$, $d = .24$), and \textit{line} ($t(1255) = 5.86$, $p < .001$, $d =.33$). For \textit{sharp\_fast}, it was also significant for \textit{curve\_fast} ($t(1255) = 3.74$, $p < .01$, $d = .21$), \textit{curve\_slow} ($t(1255) = 3.59$, $p < .01$, $d = .20$), and \textit{line} ($t(1255) = 5.17$, $p < .001$, $d = .29$). Additionally, \textit{sharp\_slow} resulted in significantly lower valence than the \textit{none} condition ($M = -.09$, $SD = 1.74$, $t(1255) = 3.35$, $p = .012$, $d = .19$). Other pairwise comparisons were not significant (all $p > .05$; $t(1255)$ range: .15–2.66, $d$ range: .01–.15).

\textbf{Arousal.} Arousal was significantly influenced by both \textit{Haptic} roughness and visual \textit{Line Style}, with high roughness (\textit{RH}) conditions and \textit{sharp} outlines leading to elevated emotional activation.
Results revealed significant main effects for both \textit{Haptic} stimuli ($F_{2,1255} = 550.06, p<.001, \eta^2_{\!p} = .47 $) and \textit{Line Style} ($F_{5,1255} = 19.33, p<.001, \eta^2_{\!p} = .05 $), with no significant interaction ($F_{10,1255} = .64, p=.780, \eta^2_{\!p} = .00 $). Post-hoc pairwise $t$-tests with Bonferroni correction were conducted separately for each main factor, namely \textit{Haptic} and \textit{Line Style}. The high roughness haptic condition (\textit{RH}; $M = 1.21$, $SD =1.88$) elicited significantly higher arousal than both the low roughness  (\textit{RL}; $M = -1.83$, $SD = 1.70$, $t(1255) = 33.09$, $p < .001$, $d = 1.87$) and medium roughness(\textit{RM}; $M = -.47$, $SD = 1.80$, $t(1255) = 18.51$, $p < .001$, $d = 1.04$) conditions. Additionally, arousal in the \textit{RM} condition was significantly higher than in the \textit{RL} condition ($t(1255) = -14.58$, $p < .001$, $d = 0.82$). Regarding \textit{Line Style}, both \textit{sharp\_fast} ($M = .17$, $SD = 1.95$) and \textit{sharp\_slow} ($M = .04$, $SD = 1.92$) elicited significantly higher arousal than \textit{line} ($M = -.79$, $SD = 1.76$; \textit{sharp\_fast}: $t(1255) = -6.52$, $p < .001$, $d = .37$; \textit{sharp\_slow}: $t(1255) = -5.70$, $p < .001$, $d = .32$), \textit{none} ($M = -.89$, $SD = .14$; \textit{sharp\_fast}: $t(1255) = -7.85$, $p < .001$, $d = .44$; \textit{sharp\_slow}: $t(1255) = -7.03$, $p < .001$, $d = .40$), and \textit{curve\_fast} ($M = -.43$, $SD = 1.81$; \textit{sharp\_fast}: $t(1255) = -4.95$, $p < .001$, $d = .28$; \textit{sharp\_slow}: $t(1255) = -4.14$, $p < .001$, $d = .23$). Similarly, \textit{curve\_slow} ($M = -.29$, $SD = 1.87$) resulted in significantly higher arousal than \textit{none} ($t(1255) = 4.23$, $p < .001$, $d = .24$) but significantly lower arousal than \textit{sharp\_fast} ($t(1255) = -3.62$, $p < .001$, $d = .20$). 
The differences between the other conditions were not statistically significant (all $p > .05$; $t(1255)$ range: .82–2.90, $d$ range: .05–.16). 

\textbf{Perceived Roughness} 
Perceived roughness was significantly shaped by both \textit{Haptic} roughness and \textit{Line Style}, with sharper visual contours and stronger haptic stimuli leading to higher roughness ratings. Results showed significant main effects of \textit{Haptic} stimuli ($F_{2,1255} = 906.8$, $p<.001$, $\eta^2_{\!p} = .59 $) and \textit{Line Style} ($F_{5,1255} = 5.95$, $p<.001$, $\eta^2_{\!p} = .02 $), while the interaction was not significant ($F_{10,1255} = .90$, $p=.53$, $\eta^2_{\!p} < .00 $). To explore these main effects, post-hoc pairwise $t$-tests with Bonferroni correction were conducted independently for the \textit{Haptic} and \textit{Line Style} conditions. The high roughness haptic condition (\textit{RH}; $M = 59.4$, $SD = 19.9$) elicited significantly higher roughness ratings than both the medium roughness(\textit{RM}; $M = 37.8$, $SD = 19.4$, $t(1255) = 24.02$, $p < .001$, $d = 1.36$) and low roughness  (\textit{RL}; $M = 22.4$, $SD = 15.1$, $t(1255) = 45.14$, $p < .001$, $d = 2.55$) conditions. Additionally, the \textit{RM} condition also produced significantly higher roughness ratings than the \textit{RL} condition ($t(1255) = -21.13$, $p < .001$, $d = 1.19$). For the \textit{Line Style} conditions, both \textit{sharp\_slow} ($M = 42.3$, $SD = 24.4$) and \textit{sharp\_fast} ($M = 42.1$, $SD = 23.6$) yielded significantly higher roughness ratings than \textit{none} ($M = 38.6$, $SD = 24.2$; \textit{sharp\_slow}: $t(1255) = -4.34$, $p < .001$, $d = .25$; \textit{sharp\_fast}: $t(1255) = -4.04$, $p < .001$, $d = .23$), \textit{line} ($M = 37.4$, $SD = 23.5$; \textit{sharp\_slow}: $t(1255) = -4.77$, $p < .001$, $d = .27$; \textit{sharp\_fast}: $t(1255) = -4.47$, $p < .001$, $d = .25$), and \textit{curve\_fast} ($M = 39.2$, $SD = 23.9$; \textit{sharp\_slow}: $t(1255) = -3.58$, $p < .01$, $d = .20$; \textit{sharp\_fast}: $t(1255) = -3.28$, $p = .02$, $d = .19$). Differences between other line styles were not statistically significant (all $p > .05$; $t$ range: $-2.87$–$1.90$, $d$ range: $-.08$–$.05$).

\subsubsection{Condition II: \textit{Color} $\times$ \textit{Haptic} }
\begin{figure*}[t]
 \centering
 \includegraphics[width=\textwidth]{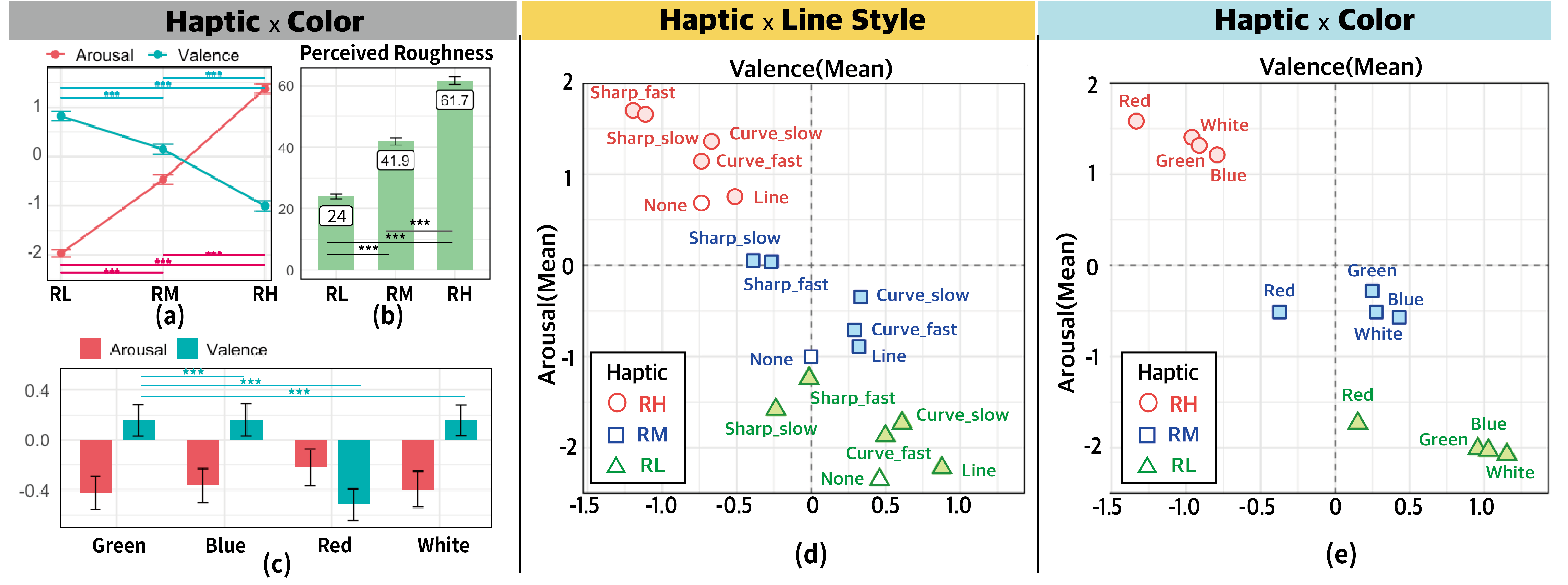}
\caption{Effects of \textit{Haptic} and \textit{Color} on affective responses and perceived roughness: 
(a) Line plot of mean arousal and valence by \textit{Haptic}. 
(b) Bar plot of perceived roughness by \textit{Haptic}. 
(c) Box plots of arousal and valence by \textit{Color}. 
(d) and (e) show scatter plots of valence and arousal across \textit{Haptic} × \textit{Line Style} and \textit{Haptic} × \textit{Color}, respectively. 
Significance: *$p<.05$, **$p<.01$, ***$p<.001$.}
\label{fig:s2colorgraph333}
\vspace{-0.4cm}
\end{figure*}

The Shapiro–Wilk test indicated significant deviations from normality (all $p < .05$) for arousal, valence, and roughness scores, prompting the use of non-parametric ART ANOVA. We summarize the key outcome measures in~\cref{fig:s2colorgraph333} (a–c), which visualizes the effects of \textit{Haptic} and \textit{Color} on affective responses and perceived roughness. 

\subsection{Results}

\textbf{Valence.} Both \textit{Haptic} roughness and \textit{Color} significantly modulated valence, with \textit{RH} stimuli and \textit{Red} leading to lower valence ratings. The results revealed significant main effects of \textit{Haptic} ($F_{2,829} =91.61$, $p < .001$, $\eta^2_p = .18$) and \textit{Color} ($F_{3,829} = 14.32$, $p < .001$, $\eta^2_p = .05$), but no significant interaction effect between the two factors ($F_{6,829} = .94$, $p = .943$, $\eta^2_p < .001$). To follow up on the main effects, pairwise $t$-tests with Bonferroni correction were independently conducted for the \textit{Haptic} and \textit{Color} conditions. The \textit{RH} \textit{Haptic} condition ($M = -1.00$, $SD = 1.76$) elicited significantly lower valence ratings than both the \textit{RL} ($M = .82$, $SD = 1.59$; $t(829) = -13.43$, $p < .001$, $d = .93$) and \textit{RM} ($M = .15$, $SD = 1.75$; $t(829) = -8.16$, $p < .001$, $d = .57$) conditions. The \textit{RL} condition also led to significantly higher valence than the \textit{RM} condition ($t(829) = 5.27$, $p < .001$, $d = .37$). For the \textit{Color} conditions, the \textit{Red} condition ($M = -.52$, $SD = 1.87$) elicited significantly lower valence than the \textit{Green} ($M = .16$, $SD = 1.83$; $t(829) = 5.37$, $d = .37$), \textit{Blue} ($M = .16$, $SD = 1.88$; $t(829) = -5.40$, $d = .37$), and \textit{White} ($M = .16$, $SD = 1.78$; $t(829) = -5.28$, $d = .37$) conditions, all $p < .001$. Differences among the other \textit{Color} pairs were not statistically significant (all $p > .05$; $t(829)$ = .02–.11, $d$ = .00–.01).

\textbf{Arousal.} Arousal was strongly modulated by haptic roughness levels, whereas \textit{Color} showed no significant effect. Results showed significant main effect of \textit{Haptic} on arousal was observed ($F_{2, 829} = 515.30, p < .001$, $\eta^2_{\!p} < .55 $), while the main effect of \textit{Color} ($F_{3,829} = 1.7$, $p = .165$, $\eta^2_{\!p} < .00 $) and no significant interaction was found between \textit{Haptic}  and \textit{Color} conditions ($F_{6,829} = 1.02$, $p = .411$, $\eta^2_{\!p} < .00 $). To further explore the main effect of \textit{Haptic}, pairwise $t$-tests with Bonferroni correction were conducted among its levels. Post-hoc pairwise comparisons with Bonferroni correction showed that the \textit{RH} condition ($M = 1.38$, $SD = 1.56$) elicited significantly higher arousal than both the \textit{RM} ($M = -.47$, $SD = 1.63$; $t(829) = 17.98$, $p < .001$, $d = 1.25$) and the\textit{RL} condition ($M = -1.96$, $SD = 1.39$; $t(829) = 32.02$, $p < .001$, $d = 2.23$). Additionally, the \textit{RM} condition yielded significantly higher arousal than the \textit{RL} condition ($t(829) = -14.05$, $p < .001$, $d = .97$). No significant differences were observed between the \textit{Color} conditions (all $p > .05$; $t$: .03–1.96, $d$: .02–.14).

\textbf{Perceived Roughness.} Perceived roughness was robustly modulated by haptic feedback, while \textit{Color} had no discernible effect. A significant main effect of \textit{Haptic} on perceived roughness was found ($F_{2,829} = 534.78$, $p < .001$, $\eta^2_{\!p} = .62$), while neither the main effect of \textit{Color} ($F_{3,829} = 2.06$, $p = .104$, $\eta^2_{\!p} < .01$) nor the interaction effect ($F_{6,829} = .59$, $p = .736$, $\eta^2_{\!p} < .01$) was significant. Bonferroni-adjusted pairwise $t$-tests were subsequently used to compare the levels of the \textit{Haptic} condition. The \textit{RH} condition ($M = 61.7$, $SD = 20.8$) elicited significantly higher roughness ratings than both the \textit{RM} ($M = 41.9$, $SD = 20.2$; $t(829) = 16.91$, $p < .001$, $d = 1.17$) and \textit{RL} ($M = 24.0$, $SD = 14.0$; $t(829) = 35.78$, $p < .001$, $d = 2.48$) conditions. Additionally, the \textit{RM} condition yielded significantly higher ratings than the \textit{RL} condition ($t(829) = -18.87$, $p < .001$, $d = 1.31$). Differences between color conditions were not statistically significant (all $p > .05$; $t(829)$ range = .34–2.43, $d$ range = .02–.17).

\section{Discussion}

Building on the key findings, we further discuss how the independent effects of visual and haptic modalities can inform affective interaction design in immersive environments.

\vspace{0.5em}
\textbf{Independent Contributions of Visual and Haptic Feedback.}
The absence of interaction effects in both \textit{Haptic $\times$ Line Style} and \textit{Haptic × Color} conditions indicates that visual and haptic feedback operated independently in shaping affective experience and perceived roughness. \added{Sensory modalities do not always integrate; instead, they are weighted by contextual reliability~\cite{ernst2004merging}. For example, task-relevant or more salient inputs tend to dominate~\cite{badde2020modality}. In our case, strong haptic cues may reduced reliance on visual cues, explaining the observed independence. This modality separation supports flexible design, allowing context-aware modulation of each channel. }
Haptic feedback, particularly roughness intensity, consistently modulated arousal and roughness perception while also influencing valence. Meanwhile, visual \textit{Line Style} significantly affected all three measures (valence, arousal, and roughness), and color influenced valence only. This suggests that visual and haptic cues acted additively rather than synergistically.

From a design perspective, this independence provides flexibility: developers can selectively target emotional or perceptual outcomes by manipulating either the visual or haptic channel without requiring multimodal congruency. For example, rough haptic stimuli can be used to reliably increase arousal regardless of visual feedback, while soft or curved visuals can be used to enhance valence.

\textbf{Visual Amplification under Limited Haptic Output.} 
\added{Although statistical interaction effects were absent, compensatory patterns under low haptic intensity suggest that visual–haptic interplay may emerge under certain conditions. This may reflect modality dominance, where participants prioritized strong and reliable haptic cues. In contrast, when haptic input was weaker, visual feedback played a greater compensatory role. This aligns with prior research on sensory reliability and crossmodal weighting}\textcolor{blue}{~\cite{ badde2020modality, ernst2004merging}}.   As illustrated in~\cref{fig:s2colorgraph333} (d–e), visual feedback such as sharp contours and red color extends the range of valence responses especially when paired with low or medium intensity haptic stimuli. For instance, in the \textit{Haptic × Line Style} condition, sharp contours at the medium haptic level reached valence zones overlapping higher-intensity haptic effects. Similarly, curved and line styles maintained more positive valence across a wider range of tactile intensities. This suggests that when haptic output is weak, visual feedback compensates by amplifying or sustaining affective modulation.

In the \textit{Haptic $\times$ Color} condition, red was the only color to significantly lower valence beyond the mid-range cluster formed by green, blue, and white, again highlighting how visual cues can diversify emotional tone when tactile input alone is insufficient. These results demonstrate that visual feedback is particularly valuable in moderate or constrained tactile conditions, enabling designers to shape valence outcomes more flexibly through visual augmentation. Such crossmodal balancing can be leveraged to maintain emotional expressiveness in mobile, lightweight, or power-limited VR systems.

\textbf{Not All Visual Cues Are Equal: Line Style vs. Color.}
The results show that \textit{line style} had a broader and more robust impact than \textit{color}. Curved and line styles were associated with higher valence, while sharp styles led to higher arousal and roughness perception. In contrast, color influenced only valence, with no significant effects on arousal or perceived roughness. These findings highlight the expressive potential of visual contour design for affective and tactile modulation in virtual interaction. Designers may consider prioritizing line-based hand feedback for richer multimodal control, while using color primarily as a valence-oriented adjustment layer.

\textbf{Modality-Specific Sensitivity.}  
Our findings suggest that users are more sensitive to affective and perceptual changes introduced by haptic stimuli than visual ones. While vibrotactile feedback significantly influenced valence, arousal, and perceived roughness, visual feedback exhibited differentiated effects depending on its form. This modality-specific responsiveness suggests prioritizing haptic feedback when designing emotionally or perceptually rich experiences, with visual cues serving a complementary or expressive role.

\subsection{Design Implications for Affective Feedback in Immersive Interaction}
\added{This work presents an initial design approach that positions the avatar as an expressive medium to expand affective feedback in immersive multimodal interfaces. Our study moves beyond single-modality feedback by examining how visuo-haptic cues modulate affective roughness perception. While Dozio et al.~\cite{dozio2022design} highlight the emotional role of multisensory cues in VR, they do not address how avatars can be intentionally designed as affective channels. Our approach uses hand-centered visual feedback to expand the affective design space, framing the avatar as a dynamic emotional interface during interaction.}

Based on the results, we propose several design considerations for integrating affective visual cues with tactile feedback in immersive interfaces. Affective feedback is crucial for shaping users’ emotional interpretation of tactile interactions in immersive environments. In emotionally sensitive or narrative-driven scenarios, the ability to flexibly modulate valence and arousal is critical for designing adaptive interfaces. \replaced{Our findings suggest that line styles have a broader influence on perceived roughness, arousal and valence than color.}{Our findings suggest that visual elements—particularly line styles—exert a broader influence on perceived roughness, arousal, and valence than color alone.} Each visual factor evokes distinct emotional outcomes, making careful selection essential for intended affective modulation.

Line styles such as \textit{sharp} significantly increase both arousal and perceived roughness, making them effective for scenarios requiring urgency or heightened attention. In contrast, \textit{line} or \textit{curve} styles promote higher valence and lower arousal, supporting calmer interactions. Color serves primarily as a valence modulator; red consistently reduced valence and is best reserved for signaling tension or alert, while blue, green, and white are better suited for inducing positive emotional tones. 

\replaced{Haptic intensity is a key driver of arousal, with stronger stimuli evoking higher levels and providing a primary means of emotional control.}{Haptic feedback intensity is the dominant driver of arousal. Stronger vibrotactile stimuli evoke higher arousal levels, offering a primary control mechanism for emotional intensity.} However, when haptic output is limited, visual contours can compensate by conveying texture or amplifying emotional tone. For instance, pairing low-intensity haptics with sharp visuals can effectively expand the emotional range of interaction.

In dynamic contexts such as gaming or training simulations, designers can manipulate both haptic intensity and visual contour to match varying emotional demands. Sharp visuals combined with high-intensity feedback are effective for conveying stress or urgency, whereas smooth visuals and low-intensity haptics are ideal for therapeutic or relaxation content. In emotionally neutral or flexible contexts, curve-based styles offer a balanced approach, slightly raising arousal while maintaining positive valence.

Overall, these results support a multi-layered feedback strategy; use haptics to broadly control arousal, and visual cues to fine-tune valence within that range. This approach enables adaptive emotional interaction under hardware constraints and supports expressive design in various VR applications, such as product exploration, education, simulation, and therapy.
\begin{figure*}[t]
 \centering
 \includegraphics[width=\textwidth]{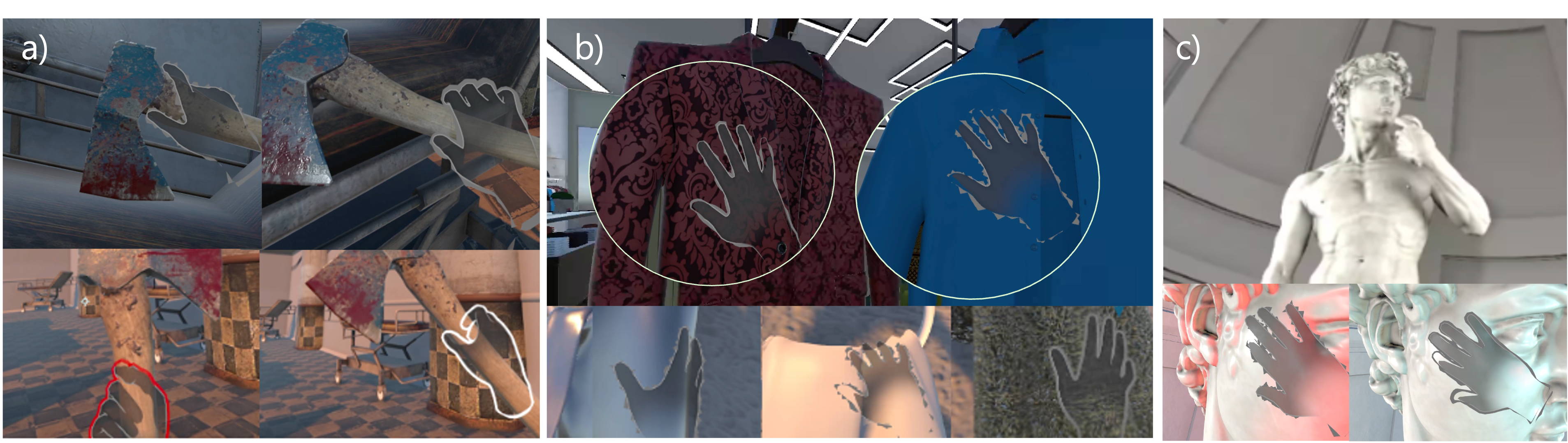}
\caption{
Application examples of visual feedback for affective modulation: 
(a) Grasping objects in horror VR using sharp contours and red tones to enhance tension; 
(b) Adjusting hand outlines to convey tactile qualities and emotional cues; 
(c) Context-driven visual feedback design (e.g., horror vs. calming environments).
}    \label{fig:line_color_va_roughness}
\label{fig:sapp}
\vspace{-0.3cm}
\end{figure*}
\subsection{Application Scenarios}
In terms of practical applications, these findings offer valuable guidance for designing emotionally adaptive experiences across diverse immersive contexts. In \textit{therapeutic VR}, smooth visual contours (such as curves or lines), combined with soft haptic feedback, can reduce anxiety and promote emotional comfort. Conversely, in \textit{training and simulation systems}, sharp visual feedback paired with high-intensity vibrations can simulate urgency, enhance realism, and support decision-making in high-stakes environments.

In \textit{gaming and horror VR}, as illustrated in ~\cref{fig:sapp} (a), line styles and color can be used to evoke either positive or negative affect when grasping virtual objects. Sharp contours and intense haptic cues heighten arousal and tension, deepening the sense of immersion. For \textit{virtual product interaction}, as shown in ~\cref{fig:sapp} (b), designers can modulate perceived material roughness and affective tone within the same arousal range using visual contours. This helps users form emotional impressions of material properties during digital prototyping or online shopping. Additionally, as shown in~\cref{fig:sapp} (c), designers can tailor visual styles to match the content’s context. For example, using sharp, red visuals to evoke fear, while other visual combinations can support calming effects. These combinations allow for intentional modulation of arousal and valence in line with the narrative or experiential goals.

Importantly, visual expression is not always beneficial. Visual feedback tends to elevate arousal, which may not be desirable in every scenario. For content aimed at relaxation or low-stimulus interaction, minimizing visual feedback or removing it altogether can help suppress unnecessary arousal. Alternatively, using \textit{curve} styles may slightly raise arousal without disrupting immersion, offering a gentle emotional lift. These examples demonstrate how combining visual and tactile design cues allows designers to shape emotional tone and perceptual clarity in context-specific and user-sensitive ways.

\subsection{Limitations and Future Work}
While this study demonstrates the independent contributions of visual and haptic cues to affective roughness perception, several directions remain for future exploration. First, \replaced{we used only the \textit{Line} contour to isolate color effects, as it did not differ significantly from the \textit{None} condition. This limited the exploration of color–contour interactions. Future work would combine color with other contour types to gain a fuller understanding of multimodal affect.}{we applied color only to the \textit{Line} style to isolate its effect; although this design was validated by the non-significant difference between \textit{None} and \textit{Line}, future work should explore how color interacts with other visual contours.} 

In addition, we focused on a limited set of visual (line shape, color) and haptic (intensity, duration) parameters. Future work may broaden the design space by incorporating other affective cues such as motion, brightness, frequency, or texture (e.g., softness). The spatial placement of visual cues (e.g., on hand vs. object) and the mapping of realistic material textures may also impact affective interpretation. 

\added{Although no interaction effects were found, future work should investigate whether such effects emerge under different contexts or stimulus conditions.}
\replaced{ We also did not consider individual or cultural differences. Future studies should include diverse participants to support culturally adaptive designs.}{Third, individual differences in visual or tactile sensitivity and cultural factors were not accounted for in this study. Future studies should include more diverse participant populations to support culturally adaptive feedback systems.} 

\added{Lastly, although we focused on the emotional effects of color (e.g., valence modulation), color can also convey contextual meanings, such as temperature (red/blue) or safety cues (red/green). Future research should examine how such symbolic meanings interact with affective cues. Additionally, the roles of embodiment and dominance remain underexplored; studying how visuo-haptic feedback affects body ownership and control could yield further insights.} Evaluating this approach in ecologically valid VR contexts such as training, therapy, or gaming may reveal broader design implications for affective interfaces.

\section{Conclusion}

This study proposes a method to modulate emotional roughness perception by combining fingertip vibrotactile patterns with visual feedback, such as \textit{Line Style} and \textit{color}. We examine how visual hand cues influence affective roughness perception in VR. Results show that visual and tactile modalities independently modulate user experience, with no observed interaction effects. For visual effects, \textit{Line Style} influenced valence, arousal, and perceived roughness, with sharp styles increasing arousal and roughness while lowering valence. \textit{Color} affected only valence, with \textit{Red} producing the strongest reduction. These results suggest visual feedback can be selectively tuned to target specific emotional or perceptual outcomes. Interestingly, when roughness was low, visual changes primarily modulated arousal and valence within the middle range. In contrast, under high roughness stimuli, tactile feedback became more dominant, reducing the influence of visual modulation on arousal and valence. These findings indicate that visual and tactile cues act additively, providing flexibility in affective interface design where each modality can be independently tuned. Even with constrained tactile feedback, visual–tactile integration supports richer emotional experiences and expands design potential for immersive content.

\acknowledgments{%
This work was partly supported by the Institute of Information \& Communications Technology Planning \& Evaluation(IITP)-ITRC(Information Technology Research Center) grant funded by the Korea government(MSIT)(IITP-2025-RS-2024-00436398) and Institute of Information \& communications Technology Planning \& Evaluation (IITP) grant funded by the Korea government (MSIT) (No. RS-2024-00397663, Real-time XR Interface Technology Development for Environmental Adaptation) and Institute of Information \& communications Technology Planning \& Evaluation (IITP) under the metaverse support program to nurture the best talents (IITP-2025-RS-2022-00156435) grant funded by the Korea government(MSIT).%
}

\bibliographystyle{abbrv-doi-hyperref}

\bibliography{template}

\appendix 

\end{document}